\documentclass[aps,pra,epsf,superscriptaddress,amsmath,amssymb,amsfonts,twocolumn,showpacs,nofootinbib,longbibliography]{revtex4-1}

\usepackage{graphicx}
\usepackage{dcolumn}
\usepackage{bm}
\usepackage{braket}
\usepackage{amsmath}
\usepackage{mathtools}
\usepackage{graphicx,color,xcolor}
\usepackage{hyperref}
\usepackage[capitalize]{cleveref}
\newcommand{\abs}[1]{\left| #1 \right|} 
\usepackage{multirow}

\usepackage{amssymb}
\usepackage{eqnarray,amsmath}
\usepackage{dcolumn}
\usepackage{graphicx}
\usepackage{natbib}
\usepackage{bbm}
\usepackage{amssymb,amsmath}
\usepackage{latexsym}
\usepackage{amsfonts}
\usepackage{epsfig}
\usepackage{psfrag}
\usepackage{graphicx}
\usepackage{hyperref}
\usepackage{color}
\usepackage{soul,xcolor}
\usepackage[normalem]{ulem}
\usepackage[titletoc]{appendix}
\usepackage{float}
\usepackage{dcolumn}
\usepackage{bm}
\usepackage{braket}
\usepackage{amsmath}
\usepackage{physics}
\usepackage{mathtools}
\usepackage{verbatim}
\usepackage{graphicx,color,xcolor,colortbl}

\usepackage[normalem]{ulem} 

\newcommand{\corr}[1]{{\color{blue} #1}}

\begin{document}
\title{Generic transverse stability of kink structures in atomic and optical nonlinear media with  competing attractive and repulsive interactions}

\author{S. I. Mistakidis}
\affiliation{Department of Physics, Missouri University of Science and Technology, Rolla, MO 65409, USA}

\author{G. Bougas}
\affiliation{Department of Physics, Missouri University of Science and Technology, Rolla, MO 65409, USA}

\author{G. C. Katsimiga}
\affiliation{Department of Physics, Missouri University of Science and Technology, Rolla, MO 65409, USA}%

\author{P. G. Kevrekidis}
\affiliation{Department of Mathematics and Statistics, University of Massachusetts Amherst, Amherst, MA 01003-4515, USA}%
\affiliation{Department of Physics, University of Massachusetts Amherst, Amherst, Massachusetts, 01003, USA}

\date{\today}

\begin{abstract} 

We demonstrate the existence and 
stability of one-dimensional (1D) topological kink configurations immersed in higher-dimensional bosonic gases and nonlinear optical setups. 
Our analysis pertains, in particular, to the two- and three-dimensional extended Gross-Pitaevskii models with quantum fluctuations  describing droplet-bearing environments but also to the two-dimensional cubic-quintic nonlinear Schr\"odinger equation containing higher-order corrections to the nonlinear refractive index. 
Contrary to the generic dark soliton transverse instability,
the kink structures are {\it generically robust} under the interplay of low-amplitude attractive and high-amplitude repulsive interactions. 
A quasi-1D effective potential picture dictates the existence of these defects, while their stability is obtained numerically and analytically through 
linearization analysis and direct dynamics in the presence of external fluctuations showcasing  their unprecedented resilience. 
These {\it generic} (across different models) findings should be detectable in current cold atom and optics experiments, offering insights towards controlling topological  excitations. 

\end{abstract}

\maketitle

\paragraph*{\textit {Introduction.}} \label{sec:intro}
Kink solitons (alias domain walls) are nonlinear excitations encountered in disparate disciplines ranging from optical~\cite{kartashov_surface_2006}, and magnetic media~\cite{perez_crossed_2008,Buijnsters_motion_2014}, living cellular structures~\cite{Ivancevic_sine_2013,Yomosa_soliton_1983,caspi_toy_1999}, folding protein chains~\cite{hu2011discrete,hu2011towards} to atomic gases~\cite{tylutki2020collective,katsimiga2023interactions} and cosmology~\cite{Vilenkin_cosmic_2000}.
Their topological character has been recently unveiled in two-dimensional (2D) van der Waals materials~\cite{lee_mobile_2023}, opening up the possibility of robust computations~\cite{wang_two_2022}.  
Stabilization of multidimensional spatially localized states is a fundamental challenge of scientific interest in physics and beyond~\cite{kartashov2019frontiers}.
It is well-known that 1D topological (e.g. dark solitons) and non-topological (bright solitons) defects suffer from the so-called snake-instability~\cite{tikhonenko1996observation,mamaev1996propagation,KIVSHAR2000117,anderson2001watching} and infrared catastrophe~\cite{fibich2015nonlinear}, respectively, once embedded in 2D and three-dimensional (3D) geometries. 
Mechanisms of suppression of the ensuing instabilities of defects have also been discussed.
These mainly consider nonuniform media, i.e., incorporating external trapping geometries~\cite{proukakis_analogies_2004,wen_dark_2013,Kevrekidis_avoiding_2004}, or are accompanied by nonlocal interactions~\cite{PhysRevE.73.066603,Lin:08,nath2008stability}, or account for
fractional dispersion in the presence of competing 
non-linearities~\cite{zeng2020preventing}.

Cold atoms are ideal quantum many-body simulators, i.e., platforms
featuring remarkable tunability of system parameters, such as e.g. interactions and external traps~\cite{bloch2008many,lewenstein2012ultracold}. 
In this context, ultradilute and incompressible self-bound states of matter, referred to as quantum droplets~\cite{luo2021new,mistakidis2023few} were recently experimentally detected in Bose mixtures~\cite{cheiney2018bright,semeghini2018self,cabrera2018quantum,d2019observation,burchianti2020dual} and dipolar gases~\cite{chomaz2022dipolar,bottcher2020new}. 
Stabilization of these states is achieved through the counterbalance of attractive and repulsive interactions modeled by mean-field nonlinear couplings and quantum fluctuations. The latter are incorporated perturbatively through the famous Lee-Huang-Yang (LHY)~\cite{lee1957eigenvalues,larsen1963binary} correction whose sign and form depend on dimensionality~\cite{Ilg_crossover_2018} leading to an extended Gross-Pitaevskii equation (eGPE)~\cite{petrov2015quantum,petrov2016ultradilute}.
Such environments sustain also different kinds of nonlinear excitations such as dark solitons~\cite{edmonds2023dark,katsimiga2023solitary,saqlain2023dragging}, bubbles~\cite{katsimiga2023interactions,edmonds2023dark}, kinks~\cite{kartashov2022spinor,tylutki2020collective,katsimiga2023interactions} and vortices~\cite{li2018two,yougurt2023vortex}.  

From a complementary nonlinear optics-seeded perspective, 
the Cubic-Quintic (CQ) model is an extension of the famous 
nonlinear Schr{\"o}dinger model~\cite{hiro}
constituing a universal mathematical toolbox involving higher-order
nonlinearities. 
In the optical setting, CQ is utilized to study the propagation of electromagnetic waves~\cite{Filho_robust_2013,Reyna_observation_2020} in photorefractive materials.
In this context, the quintic term accounts for higher-order corrections to the nonlinear refractive index~\cite{akhmediev1995novel}, i.e., incorporating susceptibilities up to fifth-order. Experimentally, 
it is often
also relevant to include the nontrivial (dissipative) absorption effects.
The broad applicability of CQ manifests by the fact that it has been deployed to describe various phenomena, for instance, liquid waveguides~\cite{Filho_robust_2013}, special types of glasses~\cite{boudebs2003experimental}, and colloids containing metallic nanoparticles~\cite{falcao2007high}. 
In atomic gases, the cubic (quintic) coupling is related to the two- (three-) body interactions~\cite{luckins2018bose,cardoso2011one,Bulgac_dilute_2002}
(see also, e.g.,~\cite{PhysRevLett.85.1146}).

In this Letter we explore the dynamical transverse stability of kinks in a wide range of models from nonlinear optics and atomic physics,
such as the 2D and 3D eGPE settings, as well as the 2D CQ model.
A key feature of these setups is the  competition between attractive 
nonlinearities (dominant at low density) and repulsive ones
(prevailing at high density) whose presence is proved crucial for the stability of kinks. 
Specifically, the stability and robustness of these entities is demonstrated in a threefold manner. Firstly, we analytically extract and numerically evaluate the relevant for each model Bogoliubov-de-Gennes (BdG) spectrum. 
Secondly, an analytical argument supporting kink's spectral stability against transverse modulations is constructed. 
Finally, we subject the relevant one-dimensional (1D) kink solutions to external, spatially distributed fluctuations (customarily present in experimental settings) and subsequently monitor
their remarkable persistence for evolution times of the order of seconds.  This demonstrates the relevance of these 1D topological 
defects in current state-of-the-art cold atom and nonlinear optics experiments and
their structural robustness as potential information carriers. 
The topological character of the structures
is evident in that no continuous, finite energy deformation can
lead to their disappearance, and that accordingly they
preserve a form of topological charge invariant, given their
distinct asymptotics at $\pm \infty$; see, e.g.,~\cite{manton2004topological,Coleman1977} for details.

\paragraph*{\textit {Models and Theoretical Analysis.}}\label{sec:Eff_pot} 

To highlight the universal features of the existence and stability of kink solutions in higher spatial  dimensions, {\it three} different nonlinear models characterized by competing  attractive 
(at low density) and repulsive (at high density) 
interactions  are investigated.  
These correspond to the eGPEs describing, for instance, self-bound quantum droplets in 2D~\cite{petrov2016ultradilute,luo2021new} and 3D~\cite{petrov2015quantum,luo2021new}, and the CQ equation. 
Experimentally, the droplet bearing systems can be emulated e.g. by considering two hyperfine states of $^{39}$K~\cite{semeghini2018self}. 
In optics the CQ model is typically used to monitor, for instance, optical beam profiles of topological defects in liquid $\rm{CS}_2$~\cite{Reyna_observation_2020}. 
In dimensionless form (see also supplemental material (SM)~\cite{supp}) these equations read
\begin{subequations}
\begin{gather}
i \partial_t \Psi = -\frac{1}{2} \left(  \partial^2_x  + \partial^2_y    \right) \Psi + g \abs{\Psi}^2 \Psi \ln   \left (  \abs{\Psi}^2  \right), \label{Eq:2D_eGPE}   \\
i \partial_t \Psi = -\frac{1}{2} \left(  \partial^2_x  + \partial^2_y   \right) \Psi -\abs{\Psi}^2 \Psi  + g_{CQ} \abs{\Psi}^4 \Psi,  \label{Eq:2D_CQ}  \\ 
i \partial_t \Psi = -\frac{1}{2} \left(  \partial^2_x  + \partial^2_y  +  \partial^2_z   \right) \Psi +g_1\abs{\Psi}^2 \Psi  +  \abs{\Psi}^3 \Psi. \label{Eq:3D_eGPE}
\end{gather}
\end{subequations}
Here, Eqs.~\eqref{Eq:2D_eGPE} and~\eqref{Eq:3D_eGPE} refer to the 2D and 3D  eGPEs, while Eq.~\eqref{Eq:2D_CQ} is the 2D CQ model with $g_{CQ}>0$. 
The right hand side always contains the Laplacian and a
nonlinear term where the latter can be written as $F(\Psi,\Psi^{\star})$, using
$^{\star}$ to denote complex conjugation. 
In the 2D eGPE, the logarithmic nonlinearity encompasses both mean-field and first-order quantum fluctuation effects~\cite{petrov2016ultradilute,luo2021new} of strength $g>0$, while its 3D counterpart features a cubic (resp. quartic) attractive with $g_1 <0$ (resp. repulsive) mean-field (LHY) contribution~\cite{petrov2015quantum}, see also SM~\cite{supp}. 
Furthermore, in the CQ model there is an attractive (repulsive) cubic (quintic) nonlinear coupling commonly accounting for two- (three-) body interactions in the realm of cold gases~\cite{luckins2018bose,Bulgac_dilute_2002} or 
higher order corrections to the Kerr effect in nonlinear optics~\cite{akhmediev1995novel,Reyna_observation_2020}.

\begin{figure}[t!]
\centering
\includegraphics[width=\columnwidth]{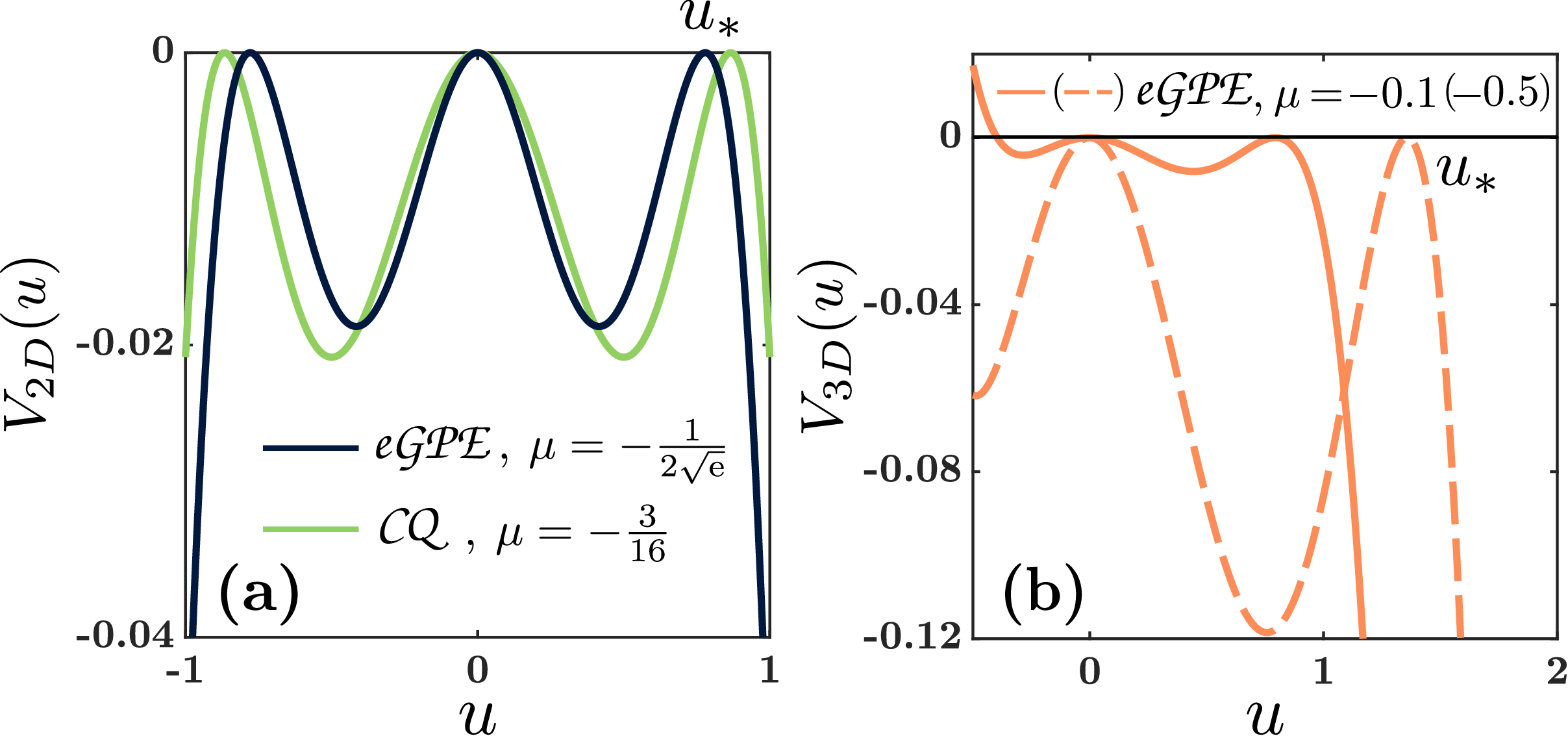}
\caption{\textit{Existence of kink configurations.} Effective potential for the three nonlinear models: (a) 2D eGPE and CQ models with $g=g_{CQ}=1$ as well as (b) the 3D eGPE setup with $g_1=-0.952$ ($g_{1}=-1.629$) for $\mu=-0.1$ ($\mu=-0.5$). The effective potentials host the kink configuration, whose wave function is constrained within $0$ and $u_*$. These states occur at specific chemical potentials (see legends).}
\label{Fig:Effective_potential}
\end{figure}

To infer the existence of planar kinks (i.e., one dimensional
ones embedded in higher dimensions), the following ansatz is employed, $\Psi(x,r_{\perp},t)=e^{-i \mu t} u(x) $. Here, $\mu$ denotes the chemical potential and $r_{\perp}=y$ ($r_{\perp}=(y,z)$) are the transverse coordinates in 2D (3D). 
A uniform background is assumed along the transverse directions, and the $u(x)$ waveform is taken to be real without loss of generality. 
Introducing this ansatz into Eqs.~\eqref{Eq:2D_eGPE}-\eqref{Eq:3D_eGPE} results in their reduced effective 1D time-independent standing wave analogues. 
These can be cast into Newtonian equations of motion, $d^2u(x)/dx^2=-dV(u)/du$, subjected to the relevant for each model effective potential $V(u)$. These turn out to be 
\begin{subequations}
\begin{gather}
V_{2D}(u) = \mu u^2 - g\frac{u^4}{2} \ln  \left(  \frac{u^2}{\sqrt{e}} \label{Eq:Pot_2D_eGPE}  \right), \\
V_{2D}(u) = \mu u^2 +\frac{u^4}{2} -g_{CQ} \frac{u^6}{3},  \label{Eq:Pot_CQ} \\
V_{3D}(u) = \mu u^2 - \frac{g_1}{2} u^4 -\frac{2}{5} u^5.  \label{Eq:Pot_3D_eGPE}
\end{gather}
\end{subequations}
Integration of the Newtonian equations of motion yields the 
effective energy, $E$, of the 1D reduced system~\cite{katsimiga2023solitary}.
While the resulting potentials feature multiple maxima, 
the frequency parameter $\mu$ can be generically tuned to 
render these maxima equi-energetic, thus enabling a 
{\it genuine heteroclinic orbit}, i.e., a kink,
between $u=0$ and $u=u_{*}$ with $E=0$.

To systematically determine the finite kink backgrounds and chemical potentials, two conditions need to be satisfied, $V(u_*)=0$ and $dV/du|_{u_*}=0$, namely, conditions for the maximum at $u=u_*$ being equi-energetic with the one at $u=0$. 
The relevant solution for potentials with a competing set of interactions yields
$\mu$ and $u_{*}$ of the kink and exists provided that the
attractive and repulsive potentials grow faster than the quadratic
power of the chemical potential term within $V$, 
and that the repulsive contribution
dominates the attractive as $u \rightarrow \infty$.
For the 2D eGPE (CQ) these values lead analytically to $\mu=-g/(2\sqrt{e})$, $u_*=e^{-1/4}$ ($\mu=-3/(16g_{CQ})$ and $u_*=\sqrt{3/(4g_{CQ})}$) [Fig. \ref{Fig:Effective_potential}(a)].
The $u_*$ boundary for the 2D eGPE 
separates the parameter region of 2D droplets from that of 2D bubbles~\cite{luo2021new,petrov2016ultradilute}, analogously to
what is the case also in 1D~\cite{katsimiga2023interactions}.
For the 3D eGPE, the chemical potential and extremum are $\mu=25g_1^3/216$ and $u_*=-5g_1/6$ respectively [Fig. \ref{Fig:Effective_potential}(b)]. Notice that
while such {kink-type} states have been discussed for CQ settings previously (see, e.g.,~\cite{Jin-Liang_2006}),
they are unprecedented, to our knowledge, for the emergent
atomic theme of 2D and 3D quantum droplet settings.
Interestingly, the asymmetry of the {effective potential $V_{3D}(u)$} [Fig. \ref{Fig:Effective_potential}(b)] suggests the 
concurrent existence of a heretofore unexplored quantum droplet
(for values of $u<0$) whose analysis is deferred for future studies.

Within the effective potential framework, unique kink solutions exist for each coupling strength when competing interactions exist,
i.e., when $g_{CQ} > 0$ and $g_1 < 0$. 
Otherwise, for $g_{CQ} \leq  0$ and $g_1 \geq 0$ kink configurations do not exist as dictated by the effective potentials of Eqs.~(\ref{Eq:Pot_CQ}) and (\ref{Eq:Pot_3D_eGPE}) respectively, see also SM~\cite{supp}. 
In what follows, without loss of generality $g$ and $g_{CQ}$ are set to unity in order to inspect the stability properties of these topological defects. On the other hand, the mean-field interaction $g_1$ in the 3D eGPE is routinely tunable in experiments via Feshbach resonances, see~\cite{semeghini2018self} and SM~\cite{supp}. 
Stability of the kink can also be ensured for other values of the interaction coefficients in the 3D eGPE and 2D CQ models as long as there is an active attractive-repulsive interplay, see for details  SM~\cite{supp}.

\paragraph*{\textit {Spectral stability of kinks in higher dimensions.}}

According to the aforementioned effective potential description the 1D planar kink solution occurs for $\mu=-1/(2\sqrt{e})$, $\mu=-3/16$, and e.g. $\mu=-0.115$ where $g_1=-1$, within the 2D eGPE, 2D CQ and 3D eGPE model, respectively.
The relevant effective 1D but also the higher dimensional stationary states are 
obtained upon utilizing heteroclinic (tanh-shaped) initial guesses, bearing the correct asymptotics. 
In the CQ case, an analytical form exists, $u=\sqrt{\frac{3}{8} (1+\tanh(\frac{\sqrt{3} x}{2 \sqrt{2}}))}$~\cite{Jin-Liang_2006,birnbaum2008families}. It is then numerically confirmed that such 
1D planar waveforms satisfy the corresponding 2D and 3D 
time-independent version of Eqs.~(\ref{Eq:2D_eGPE})-(\ref{Eq:3D_eGPE}). 

\begin{figure}[t!]
\centering
\includegraphics[width=\columnwidth]{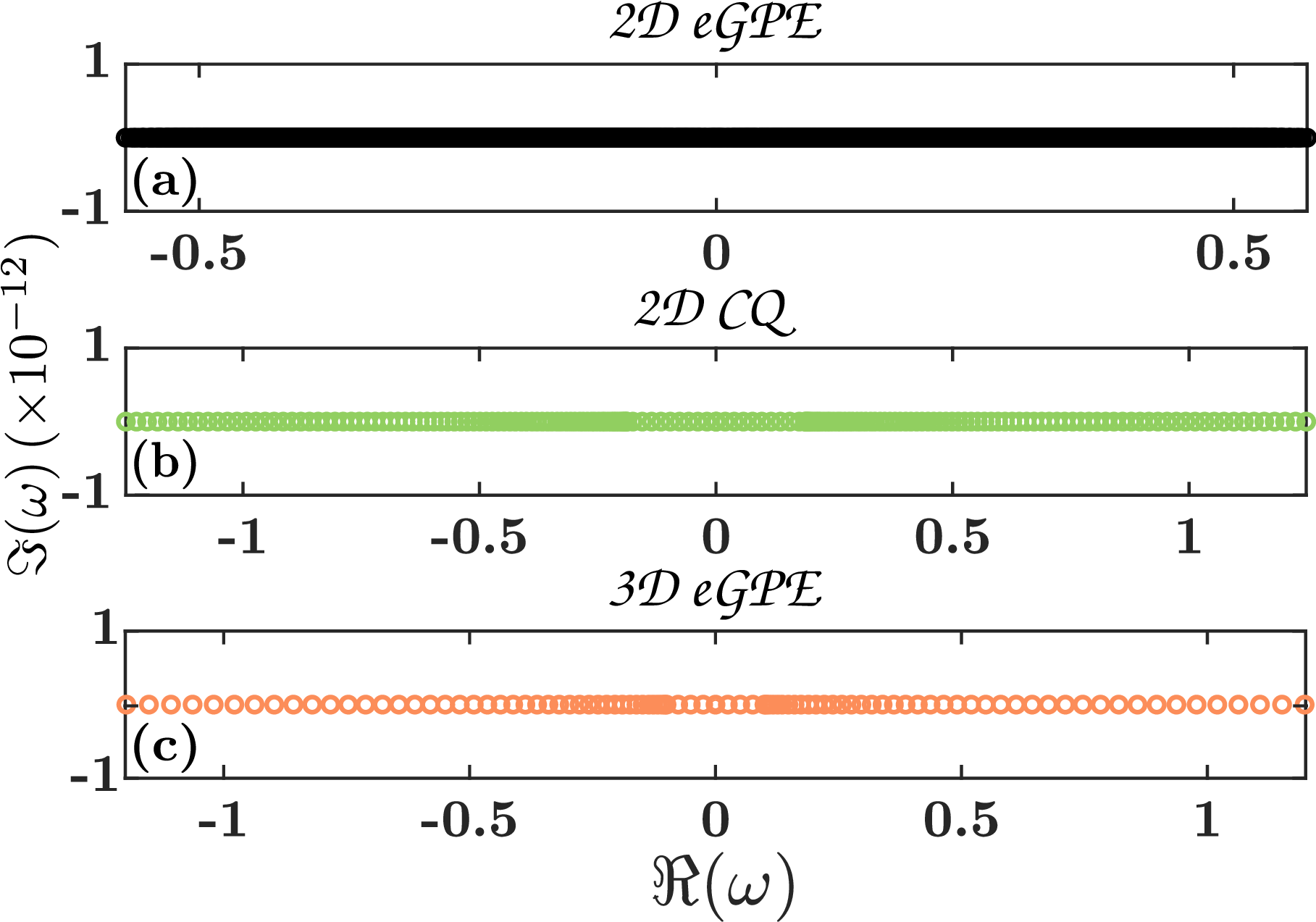}
\caption{\textit{Spectral stability of kinks.} BdG analysis 
for the kink waveforms within
(a) the 2D eGPE, (b) 2D CQ and 
(c) the 3D eGPE models.
The absence of a finite imaginary part, $\Im[\omega]\neq 0$, in the spectrum showcases the stability of the respective  kink states. }
\label{Fig:Kink_BdG_BdG}
\end{figure}

\begin{figure*}
\centering
\includegraphics[width=1\textwidth]{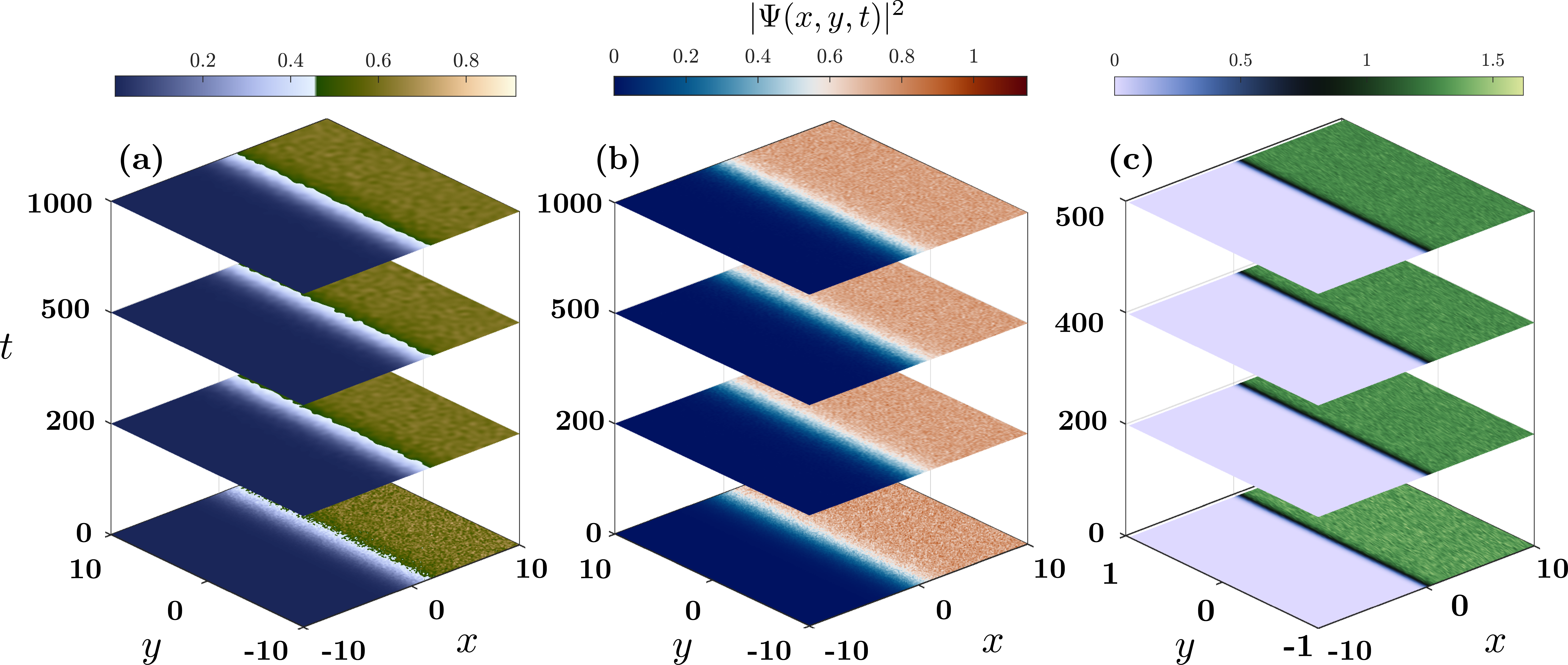}
\caption{\textit{Resilience of kink states.} Density snapshots depicting the evolution of perturbed kink solutions within the (a) 2D eGPE, (b) CQ and (c) 3D eGPE models with $\mu=-0.1$. It can be readily seen that, in all cases, the kink is robust throughout the evolution with its core remaining intact and the waveform maintaining its shape despite being significantly distorted due to external spatial and phase fluctuations. The box sizes used for the simulations correspond to $L_x=L_y=400$ and $L_z=6$.}
\label{Fig:Kink_dynamics}
\end{figure*} 

To extract the stability properties of these configurations, small perturbations of the numerically exact (up to a prescribed tolerance)
kink solutions are considered 
having the form
$\Psi= \left(\Psi_0+\epsilon a e^{-i\omega t}+\epsilon b^{\star}e^{i\omega^{\star} t}\right)e^{-i\mu t}$.   
$a, b$ ($\omega$) refer to the   
eigenvectors (eigenfrequencies), whilst $\epsilon \ll 1$ is a small perturbation parameter.
Utilizing this ansatz and keeping terms of order $\mathcal{O}(\epsilon)$ leads to an eigenvalue problem that, irrespectively of the system under consideration [Eqs.~(\ref{Eq:2D_eGPE})-(\ref{Eq:3D_eGPE})], can be written as

\begin{gather}
\begin{bmatrix}
\hat{L}_{11} & \hat{L}_{12}\\
\hat{L}_{21} & \hat{L}_{22}
\end{bmatrix}
\begin{bmatrix}
a \\
b 
\end{bmatrix}
= \omega
\begin{bmatrix}
a \\
b
\end{bmatrix}.
\label{bdg}
\end{gather}
The model-dependent diagonal matrix elements are 
$\hat{L}_{11}=-\frac{1}{2}\nabla^2-\mu+\Psi^2_0+2\Psi_0^2 \rm{ln}(\Psi^2_0)$, 
$\hat{L}_{11}=-\frac{1}{2}\nabla^2-\mu-2\Psi^2_0+3\Psi_0^4$,
and $\hat{L}_{11}=-\frac{1}{2}\nabla^2-\mu+2 g_1 \Psi_0^2+\frac{5}{2}\Psi_0 ^3$,
for the 2D eGPE, CQ, and 3D eGPE respectively.
Their relevant off-diagonal counterparts read
$ \hat{L}_{12}=\Psi^2_0+\Psi^2_0 \rm{ln}(\Psi^2_0)$,
$  \hat{L}_{12}=-\Psi^2_0+2\Psi_0^4$, 
and $ \hat{L}_{12}=g_1\Psi^2_0+\frac{3}{2}\Psi_0 ^3$, respectively.  
In all cases, $\hat{L}_{11}=-\hat{L}_{22}$, 
$\hat{L}_{12}=-\hat{L}_{21}$ and $\nabla^2$
denotes the $\mathcal{N}$-dimensional Laplace operator.
Additionally, $\Psi_0$ designates the stationary, real kink state for each distinct model.

The stability analysis outcome for all three models, is depicted in Figs.~\ref{Fig:Kink_BdG_BdG}(a)-(c). 
In all scenarios, spectral stability of the kink state can be inferred from the {\it absence} of a finite imaginary part, $\Im (\omega) \sim 10^{-12}$, in the pertinent BdG spectrum. 
For all settings, the first hundred eigenvalues of the continuous 
quasi-1D BdG spectra are illustrated, while we 
verified that their relevant higher dimensional analogues produce the same stability result. 
For the quasi-1D BdG spectra, the transverse direction is decomposed into Fourier modes with the 
``quantization'' thereof from a finite transverse computational domain
accounted for. 
The stability of a plethora of effective
1D spectra for different transverse wavenumbers is confirmed.

Moreover, an analytical argument for the kinks stability  relies on an alternative BdG formulation, utilizing the perturbation $\Psi(x,r_{\perp},t) = \left[  \tilde{u}(x) +w(x,r_{\perp},t) \right] e^{-i \mu t}$, where $w$ is a complex wave function [see also SM~\cite{supp}]. 
Here, stability is determined by the spectral operators $L_{\pm,0} = -\frac{1}{2} \partial_x^2-\mu + \frac{\partial F(u,u^{\star})}{\partial u}\lvert_{\tilde{u}} \pm \frac{\partial F(u,u^{\star})}{\partial u^{\star}}\lvert_{\tilde{u}}$ 
at transverse wavenumber $k_{\perp}=0$, where $\tilde{u}$ is the relevant kink solution. 
Direct calculation yields $L_{+,0} (\partial_x {\tilde{u}})=0$ and since $\partial_x \tilde{u}$ is nodeless, according to Sturm-Liouville theorem, it is the ground state
of $L_{+,0}$ with eigenvalue $\lambda=0$. Similarly, 
$L_{-,0} {\tilde{u}}=0$ leading ${\tilde{u}}$ to be an
eigenfunction
of the latter operator with $\lambda=0$. 
This, together with the
addition of $k_{\perp}^2/2$ for transverse modulations $ \propto e^{i k_{\perp} r_{\perp}}$
implies that the spectra of the operators $L_{\pm,k_{\perp}}=L_{\pm,0} + k_{\perp}^2/2$
are non-negative. Then, a straightforward spectral argument (provided
in SM) {\it rigorously proves that there are no exponentially growing transverse modulations of
finite $k_{\perp}$, which can destabilize the kink}, as further verified by our simulations below.

\paragraph*{\textit{Dynamical confirmation of the absence of transverse instability for the kinks.}}

To corroborate our BdG findings we next consider the dynamical evolution of the stationary kink solution for each of the models under study. 
Specifically, we aim not only to mimic unavoidable experimental imperfections but also to exemplify the robust nature of the kink structure in the presence of transverse perturbations. 
To that effect, the kink initial state is significantly perturbed via a 
random normal distribution generator, 
$\delta(x,r_{\perp})$. 
The ensuing wave function acquires the form
$\psi(x,r_{\perp},t=0)=\Psi_0 \left[1+\varepsilon \delta (x,r_{\perp})\right]$.
In this context, $\delta(x,r_{\perp})$ is characterized by zero mean and unit variance, while $\varepsilon=\varepsilon_R+i\varepsilon_I$ such that also phase disturbances are encountered. 
This implies that both the core and the background of the kink are perturbed with the latter accounting for chemical potential (particle number) fluctuations unavoidable in experiments. The same holds upon considering perturbation in the interaction coefficients (not shown for brevity). 
In particular, the considered amplitudes of the perturbation range lie in the interval $[10\%, 50\%]$ of the solution amplitude. 
Typical evolution times (in the dimensionless units adopted herein) are of the order of $\sim 
500$ and $\sim 10^3$ for the 3D and 2D settings, respectively, confirming the longevity of these planar 1D defects upon their exposure to transverse excitations.

Evidently, despite the significantly excited nature of the initial state the kink remains intact preserving its shape throughout the evolution,
while ``shedding'' some of the relevant ``radiation'' in the form
of dispersive wavepackets, as it can be seen in Figs.~\ref{Fig:Kink_dynamics}(a)-(c). Particularly here, 
distinct plates correspond to density, $|\Psi(x,y,t)|^2$, contours in the $x-y$ plane taken at different times during the propagation of the perturbed kink configuration.
For the 3D dynamics the $z$-direction is integrated out, while we note that the robustness of the kink configurations has also been verified for different values of the chemical potential and hence distinct interaction strengths.

\paragraph*{\textit {Experimental implications.}}

For droplet bearing settings, two hyperfine levels e.g. $\ket{F=1,m_F=0}$, $\ket{F=1,m_F=-1}$ of ultracold $^{39}$K in the vicinity of the $59 \, \rm{G}$ intraspecies Feshbach resonance~\cite{chin2010feshbach,Errico_Feshbach_2007} feature competing intra- and inter-species interactions~\cite{cabrera2018quantum,semeghini2018self}. 
They are experimentally populated using radiofrequency spectroscopy~\cite{cabrera2018quantum} and their dynamics can be adequately described by the eGPE models [Eqs. \eqref{Eq:2D_eGPE}, \eqref{Eq:3D_eGPE}] in 2D and 3D.
For the 3D setup with averaged mean-field scattering length $\delta a = -0.24 \, a_0$, propagation times of the kink ($\sim 500$ in dimensionless units) refer approximately to $800 \, m s$~\cite{supp}.
The corresponding 2D pancake geometry features a tight harmonic trap along one spatial dimension~\cite{Hadzibabic_two_2011} and e.g. for $\omega_z=2\pi \times 2 \, \rm{kHz}$~\cite{cabrera2018quantum,Jalm_dynamical_2019} and $\delta a =-1.5 \, a_0$ [see SM~\cite{supp}], the relevant timescale ($\sim 1000$ in dimensionless units) for the kink evolution is $2 \,$ secs. These long evolution times guarantee the resilience of the kink in current state-of-the-art experiments incurring additionally three-body recombination~\cite{semeghini2018self}. 

On the other hand, the 2D CQ equation is used to describe, for instance, propagation of electromagnetic waves in   
optical media incorporating higher-order nonlinear refractive indices~\cite{sutherland_handbook_2003}.
Characteristic materials featuring such properties are chalcogenide glasses~\cite{boudebs2003experimental}, such as 
liquid $\rm{CS}_2$~\cite{Filho_robust_2013,Reyna_observation_2020}.
In these setups (where optical vortices and their destabilization has been monitored~\cite{desyatnikov_optical_2005}), the time-evolution is probed by measuring the transmission of the kink along the axial direction of the nonlinear medium.
Considering a light beam of $900 \, nm$~\cite{Filho_robust_2013}, a 
propagation distance of $10^3$ corresponds to $0.14 \, mm$ axial depth for the kink transmission. 
Finally, in either cold bosonic mixtures or  nonlinear optical media, kink defects can be imprinted in the bulk by means of the well-established technique of density engineering~\cite{anderson2001watching,Farolfi_observation_2020,Matthews_vortices_1999,shomroni_evidence_2009}, or utilizing digital micromirror device patterned optical traps~\cite{Jalm_dynamical_2019,navon_quantum_2021,gauthier_direct_2016,tamura2023observation}. 
In the SM~\cite{supp},
we demonstrate the kink dynamical formation through
density engineering, and its potential robustness under different trap settings.

\paragraph*{\textit {Conclusions and future challenges.}}\label{conclusions} 
The existence and corresponding spectral and dynamical stability of 1D kink solutions embedded in higher-dimensional gases or nonlinear optical media featuring competing attractive lower order 
and repulsive higher order interactions is unveiled.
To establish the breadth and generality of results, 
this is exemplarily demonstrated for three representative models, the 2D and 3D eGPE droplet bearing systems and the 2D CQ setting relevant in nonlinear optics. 
We explicated the existence of these 1D topological defects upon 
analytically extracting, in each case, a quasi-1D effective potential picture, subsequently supplemented by their spectral stability inferred through linearization analysis, both numerically and analytically.  
Additionally, monitoring their dynamical evolution in the presence of external fluctuations, revealed their profound robustness and resilience, for times up to few seconds. 

These results suggest the use of kink defects as promising candidates for topological quantum computing. 
Indeed, contrary to the generically transversely unstable dark soliton states~\cite{KIVSHAR2000117}, kinks appear to {\it defy} the proneness to transverse modulations.
Importantly, also, these results pave the way for further physical and mathematical studies. 
Our effective potential analysis suggests
the existence of further lower-dimensional 
nonlinear excitations (including droplets, bubbles, etc.),
whose transverse dynamics and potential dynamical features may be
of further theoretical and experimental interest in their own right. 

\paragraph*{\textit {Acknowledgments.}}\label{Acknowledgments}
We acknowledge fruitful discussions with B. Malomed and we are grateful to J. Holmer for extensive discussions
and insights regarding the analytical treatment of the kink stability. We also thank the anonymous referee for the insightful comments. This material is based upon work supported by the U.S. National Science Foundation under the awards No. PHY-2110030, No. PHY-2408988, and No. DMS-2204702 (PGK). 
S.I.M acknowledges support from the Missouri University of Science and Technology, Department of Physics, Startup fund.

\bibliographystyle{apsrev4-1}
\bibliography{reference}	

\clearpage

\onecolumngrid
\setcounter{equation}{0}
\setcounter{figure}{0}
\setcounter{section}{0}
\makeatletter

\renewcommand{\corr}[1]{{#1}}

\newcommand{\ds}{\displaystyle}
\newcommand{\be}{\begin{equation}}
\newcommand{\ee}{\end{equation}}
\newcommand{\beq}{\begin{eqnarray}}
\newcommand{\eeq}{\end{eqnarray}}
\newcommand{\dt}{\ds\frac{\dd}{\dd t}}
\newcommand{\dz}{\ds\frac{\dd}{\dd z}}
\newcommand{\D}{\ds\left(\frac{\dd}{\dd t} + c \frac{\dd}{\dd z}\right)}

\newcommand{\hide}[1]{}
\newcommand{\app}[1]{Appx.\,(\ref{#1})}
\newcommand{\eq}[1]{Eq.\,(\ref{#1})}
\newcommand{\eqs}[1]{Eqs.\,(\ref{#1})}
\newcommand{\noeq}[1]{(\ref{#1})}
\newcommand{\fig}[1]{Fig.\,\ref{#1}}
\newcommand{\figs}[2]{Figs.\,\ref{#1} and~\ref{#2}}
\newcommand{\nofig}[1]{(\ref{#1})}
\newcommand{\tab}[1]{Table\,\ref{#1}}

\newcommand{\w}{\omega}
\newcommand{\W}{\Omega}
\newcommand{\g}{\gamma}
\newcommand{\G}{\Gamma}
\newcommand{\E}{\hat E}
\newcommand{\s}{\sigma}

\renewcommand{\theequation}{S\arabic{equation}}
\renewcommand{\thefigure}{S\arabic{figure}}
\renewcommand{\bibnumfmt}[1]{[S#1]}
\setcounter{page}1
\def\thepage{S\arabic{page}}

\begin{center}
	{\Large\bfseries Supplemental  Material: \\ 
 Generic transverse stability of kink structures in atomic and optical nonlinear media with  competing repulsive and attractive interactions}
\end{center}

\section{Supporting stability of the kink in the  droplet and CQ models}  \label{Supp:stability_additional}

\begin{figure}[h]
\centering
\includegraphics[width=\textwidth]{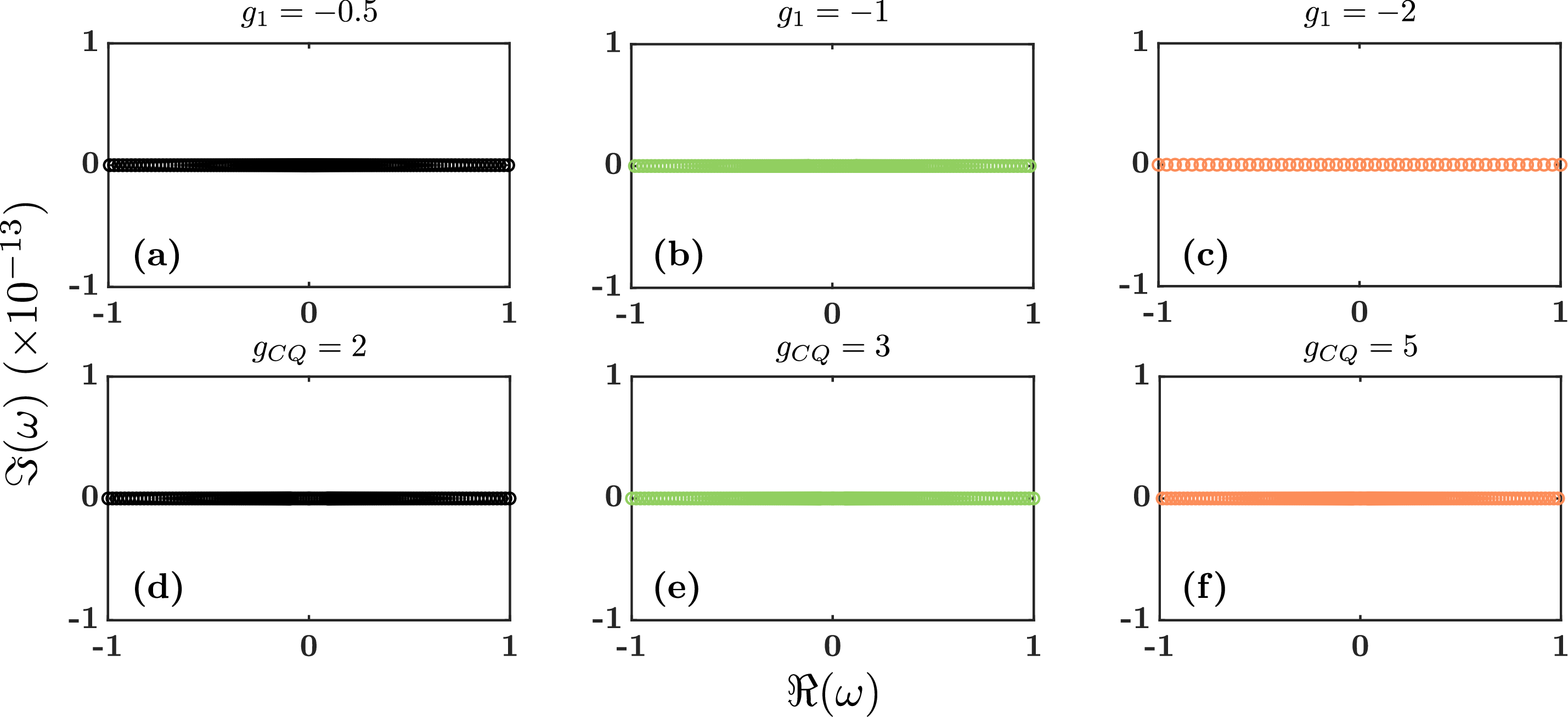}
\caption{Real part, $\Re(\omega)$, with respect to the imaginary, $\Im(\omega)$, part of the eigenfrequencies $\omega$ of the kink soliton within the (a)-(c) 3D eGPE and (d)-(f) 2D CQ model for differrent interaction coefficients (see legends). Spectral stability against transverse modes is inferred by the zero imaginary contribution in all cases further supporting the results of the main text.}
\label{Fig:BDGs}
\end{figure}

In what follows, we generalize our findings regarding the resilience of the kink solution presented in the main text by elaborating further on its stability as we vary the participating interaction parameters in both the 3D eGPE and the 2D CQ model.
Recall, that in these two setups kink waveforms exist for distinct values of the cubic ($g_1$) and the quintic ($g_{CQ}$) nonlinearity respectively,
as predicted by the relevant effective potential description of Eqs.~(2b) and (2c) discussed in the main text. 
The real, $\Re(\omega)$, vs the imaginary, $\Im(\omega)$, part of the ensuing 
eigenfrequencies is illustrated in Fig.~\ref{Fig:BDGs}(a)-(c) [Fig.~\ref{Fig:BDGs}(d)-(f)] within the 3D eGPE [2D CQ] model for progressively more attractive [repulsive] interactions i.e. $g_1=-0.5, -1, -2$ [$g_{CQ}=2, 3, 5$]. 
We remark here that in the 3D eGPE scenario increasingly large in magnitude couplings are currently more challenging to realize experimentally due to unavailable Feshbach resonances~\cite{semeghini2018self}. 
Of course, this does not preclude the possibility that they can become available in near future  experiments, and are certainly of interest from a mathematical viewpoint.  
It is important to clarify that for $g_{CQ} \leq 0$ kink solutions are absent within the CQ model in accordance with the predictions of the effective potential picture which in this case exhibits a local minimum 
at $u=0$ and hence does not support kink configurations. 
A similar argument leads to the absence of
kinks for $g_{1} \geq 0$ in the 3D eGPE setting. 
As such, it is imperative to emphasize once again that solely in the presence of competing attractive and repulsive interactions the kink waveform does not suffer from transverse destabilization. 
Indeed, in line with the results discussed in the main text, spectral stability of the kink solution is evident ($\Im(\omega)\sim 10^{-13}$) irrespectively of the system under consideration and the relative strength of competing attractive/repulsive interactions, see Fig~\ref{Fig:BDGs}. 

\section{Analytical aspects of the kink stability in all considered models} \label{Supp:Analytics}

To analytically establish kink stability across all models under consideration, we employ the following perturbation ansatz
\begin{equation}
    \Psi(x,y,t) = [\tilde{u}(x) + w(x,r_{\perp},t)] e^{-i \mu t},
    \label{Eq:Perturb_expansion}
\end{equation}
where $\tilde{u}(x)$ is the 1D profile of the kink, coinciding with the expression stemming from the reduced system of Eqs. (2a)-(2c) in the main text.
Moreover, $\mu<0$ is the corresponding chemical potential and $w$ is a complex wave function satisfying $\abs{w} \ll \tilde{u}, ~ \forall r_{\perp},t$.

Substituting the above ansatz~\eqref{Eq:Perturb_expansion} into Eqs. (1a)-(1c) in the main text, and retaining only terms linear in $w$, we obtain,
\begin{subequations}
\begin{gather}
    i \partial_t w(x,r_{\perp},t) = L_+ \Re \{w(x,r_{\perp},t)\} + i L_- \Im \{w(x,r_{\perp},t)\}, \quad \rm{where} \label{Eq:Supp_linear} \\
    L_{\pm} = -\frac{1}{2} \nabla^2 -\mu + 
    \frac{\partial F(u,u^{\star})}{\partial u} \lvert_{\tilde{u}} \pm \frac{\partial F(u,u^{\star})}{\partial u^{\star}} \lvert_{\tilde{u}}.
\end{gather}
\end{subequations}
Here, $F(u,u^{\star})$ represents the nonlinear terms added to the
Laplacian in Eqs. (1a)-(1c) in the main text. 
For simplicity, we drop the hat symbol from the operators. 
The linearized equations can be written in the following compact form,
\begin{equation}
     \partial_t \begin{bmatrix}
         \Re \{w(x,r_{\perp},t)\} \\ \Im \{w(x,r_{\perp},t)\}
     \end{bmatrix} =
     \begin{bmatrix}
         0 && L_- \\ -L_+ && 0
     \end{bmatrix}
     \begin{bmatrix}
         \Re \{w(x,r_{\perp},t)\} \\ \Im \{w(x,r_{\perp},t)\}
     \end{bmatrix} \label{Eq:Linear_matrix_format}.
\end{equation}
It is convenient to Fourier transform the perturbation along the transverse directions, $\hat{w}(x,k_{\perp},t) = \int dr_{\perp} ~ e^{-i k_{\perp} r_{\perp}} w(x,r_{\perp},t)$, since the generic destabilization mechanism of 1D states embedded in higher dimensions is via transverse excitations~\cite{Bougas_stability_2024}.
The nabla operator therefore transforms as $\nabla^2 \to \partial_x^2 -k_{\perp}^2$, and Eq.~\eqref{Eq:Linear_matrix_format} becomes effectively 1D,
\begin{subequations}
\begin{gather}
    \partial_t \begin{bmatrix}
         \Re \hat{w} \\ \Im \hat{w}
     \end{bmatrix} =
     \begin{bmatrix}
         0 && L_{-,k_{\perp}} \\ -L_{+,k_{\perp}} && 0
     \end{bmatrix}
     \begin{bmatrix}
         \Re \hat{w} \\ \Im \hat{w}
     \end{bmatrix} 
     , \label{Eq:Linear_matrix_format_mom} \\
     L_{\pm,k_{\perp}} = -\frac{1}{2} \partial_x^2 +\frac{1}{2} k_{\perp}^2 -\mu +
      \frac{\partial F(u,u^{\star})}{\partial u} \lvert_{\tilde{u}}\pm\frac{\partial F(u,u^{\star})}{\partial u^{\star}} \lvert_{\tilde{u}}. 
\end{gather}
\end{subequations}
Note that we have dropped the arguments of $\hat{w}$ for simplicity.
So far we have repeated the BdG analysis, keeping
the setup very general for quasi-1D configurations (which are
stable in 1D) under
transverse modulations. The aim is to explore whether these
transverse modulations introduced by $k_{\perp} \neq 0$ lead to
instability in the case of the kink structure.

Instead of solving however the matrix equation~\eqref{Eq:Linear_matrix_format_mom}, we focus on the eigenspectra of the $L_{\pm,k_{\perp}}$ operators. As indicated in the main text, it is possible
to identify the eigenvectors of the $L_{\pm,k_{\perp}}$ operators
with vanishing eigenvalues for the 1D case of $k_{\perp}=0$ (no transverse
modulation). In particular, it can be analytically shown,
through a generic, direct calculation
(arising from the spatial differentiation of
the steady state equation), that $L_{+,0} (\partial_x {\tilde{u}})=0$.
Due to the form of the kink soliton (tanh-like), its derivative, $\partial_x \tilde{u}$, does not possess any nodes.
Given that $L_{+,0}$ is a second-order linear differential operator, the Sturm-Liouville theory~\cite{arfken_mathematical_1972} then implies that $\partial_x \tilde{u}$ is the ground state of $L_{+,0}$ with eigenvalue $\lambda=0$.
Moreover, since the eigenvalues of $L_{+,0}$ are ordered, the rest of the spectrum (which turns out to be continuous spectrum) is positive. 
Similarly, $L_{-,0} {\tilde{u}}=0$ leading ${\tilde{u}}$ to be an
eigenfunction of the latter operator with $\lambda=0$. 
Again, $\tilde{u}$ does not possess any nodes, and Sturm-Liouville theory suggests that $\tilde{u}$ is the ground state of $L_{-,0}$ with $\lambda=0$, and the rest of the spectrum is positive. 
Importantly, \textit{there are no} $\lambda<0$ for these
operators.
In the presence of transverse modulations $ \propto e^{i k_{\perp} r_{\perp}}$
the formulae $L_{\pm,k_{\perp}}=L_{\pm,0} + k_{\perp}^2/2$
clearly indicate that the spectra of $L_{\pm,k_{\perp}}$ 
lie in the positive half-line (augmented from that of the case $k_{\perp}=0$
through the $k_{\perp}^2/2$ addition term). Now we present an argument
{\bf rigorously} establishing that for $k_{\perp} \neq 0$ (associated
with transverse modulations), this 
non-negative spectrum is tantamount to {\bf spectral stability} of
the kink structure which is stable in 1D to such transverse modulations.

To understand how the individual spectra carry information regarding the stability of the kink structures, 
we proceed as follows.
Firstly, given that $L_{\pm,k_{\perp}}$ are positive definite
operators for $k_{\perp} \neq 0$, it follows that
$\sqrt{L_{-,k_{\perp}}} L_{+,k_{\perp}} \sqrt{L_{-,k_{\perp}}}$ is also positive definite.
Considering any eigenvector of $L_{+,k_{\perp}}$ such that
$L_{+,k_{\perp}} u^{(k)}(x)= \epsilon^{(k)} u^{(k)}(x) $, with 
$\epsilon^{(k)}>0$ we can construct the ancillary vector, $g(x) = \sqrt{L_{-,k_{\perp}}}^{-1} u^{(k)}(x)$. 
Since $L_{-,k_{\perp}}$ possesses purely positive spectrum for $k_{\perp}>0$, it can be inverted.
Taking the inner product 
\begin{gather}
    \langle \sqrt{L_{-,k_{\perp}}} L_{+,k_{\perp}}   \sqrt{L_{-,k_{\perp}}} g(x), g(x) \rangle =  \langle  \sqrt{L_{-,k_{\perp}}} \epsilon^{(k)} u^{(k)}(x) , \sqrt{L_{-,k_{\perp}}}^{-1} u^{(k)}(x)   \rangle \nonumber \\
    = \epsilon^{(k)} \langle u^{(k)}(x),  u^{(k)}(x)     \rangle = \epsilon^{(k)} \norm{u^{(k)}(x)} >0,  \label{Eq:Inner_prod}
\end{gather}
where $\norm{\cdot}$ denotes the norm of a vector. 

Moreover, the operator $A=L_{+,k_{\perp}} L_{-,k_{\perp}}$ is positive definite as well, since it is related to $B=\sqrt{L_{-,k_{\perp}}} L_{+,k_{\perp}} \sqrt{L_{-,k_{\perp}}}$ via a similarity transformation provided by the positive definite $\sqrt{L_{-,k_{\perp}}}$. Therefore $A$ and $B$ share the same eigenvalues. Furthermore, $-L_{+,k_{\perp}} L_{-,k_{\perp}}$ is
 the linearization operator in Eq.~\eqref{Eq:Linear_matrix_format_mom} for the eigenvalue
$\lambda^2$, i.e. $\partial_t^2 \Im \hat{w} = -L_{+,k_{\perp}} L_{-,k_{\perp}} \Im \hat{w} = \lambda^2 \Im \hat{w}$. 
Importantly,  $\lambda^2<0$ ($\lambda^2>0$) implies the absence (existence) of exponential growth of transverse perturbations.
Since we showed that $L_{+,k_{\perp}} L_{-,k_{\perp}}$ is positive 
definite for the transverse modulation case of $k_{\perp} \neq 0$, 
this, in turn, proves that $\lambda^2 < 0$ and hence
the spectrum of the linearization operator 
is purely imaginary for the kink structure, bearing
no instability to transverse modulations.

\section{Density engineering of the kink defect}  \label{Supp:dens_engineering}

\begin{figure*}[t!]
\centering
\includegraphics[width=1\textwidth]{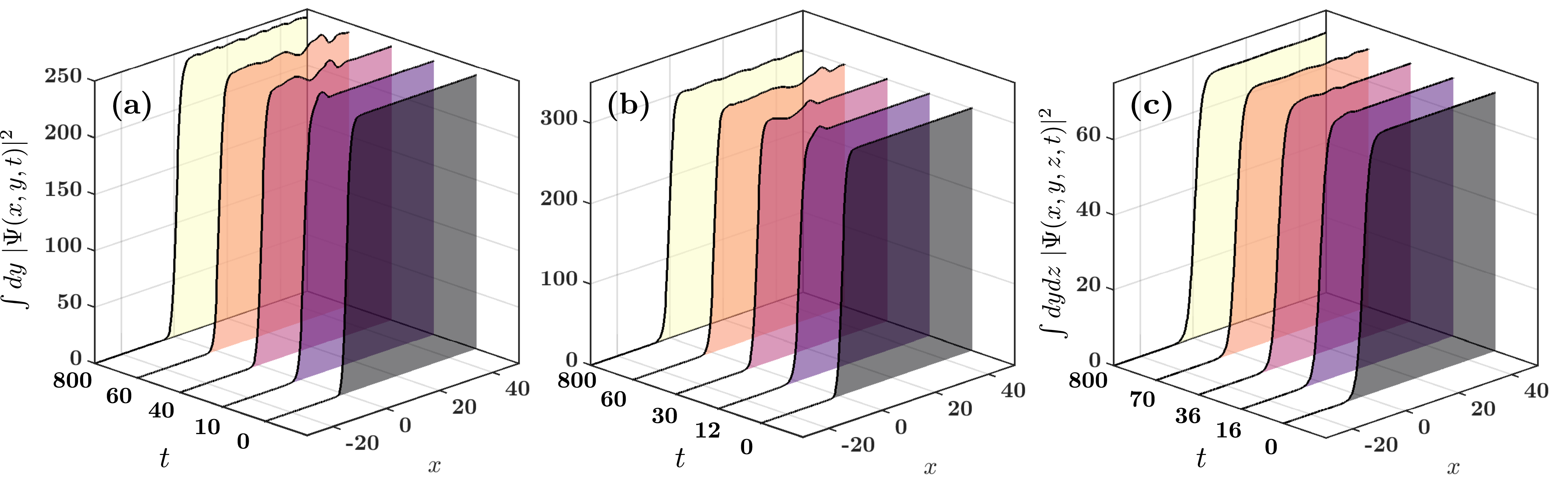}
\caption{Following the barrier release, the density engineered state (at $t=0$) transitions towards the kink configuration in the (a) 2D eGPE, (b) CQ, and (c) 3D eGPE models. Such a state is initially sculpted by employing the barrier $V(x,y)$ of Eq.~\eqref{Eq:Barrier} with $V_0 = 20$, $w=0.6$, $\frac{\sqrt{3}}{2 \sqrt{2}}$, $0.42$, and $x_0 =6$, $ 6$, $10$ in the case of the 2D eGPE, CQ, and 3D eGPE setup respectively. It is apparent that the kink remains stable independently of the model. In all panels the barrier is superimposed to a 2D box trap of length $L=400$; accordingly,
only a small portion of the domain is shown along the 
$x$-direction. In the 3D eGPE scenario (panel (c)), a harmonic trap of frequency $\omega_z = 5$ is present in the $z$ direction.}
\label{Fig:Kink_engineering}
\end{figure*}

Contemporary experiments are able to sculpt arbitrarily shaped external potentials, including box traps but also controllably imprint a plethora of different patterns in the atomic background~\cite{Jalm_dynamical_2019}.
To further corroborate the experimental realization of the kink state, we emulate the technique of density engineering~\cite{Farolfi_observation_2020,shomroni_evidence_2009,anderson2001watching}.
To do so, we deploy the following potential barrier,
\begin{equation}
    V(x,y) = \frac{V_0}{2}  \left[  1- \tanh \left(  w(x+x_0) \right) \right], 
    \label{Eq:Barrier}
\end{equation}
where $(V_0,w)$  denote the potential height and width, while $x_0$ denotes its center.
An additional external trapping potential is assumed having the form of either a 2D box for the 2D eGPE and CQ models, or a combined box across the $(x,y)$ plane along with a tight harmonic trap in the $z$ direction, i.e. $V_{trap}(z) = \frac{1}{2} \omega_z^2 z^2$, for the 3D eGPE.
We remark that the choice of the trap only along the $z$ direction is compatible with the asymptotics of the kink structure being associated with its topological nature,
as well as with our interest in exploring the
{\it transverse} (in the case along $y$) stability of the 
kink in such a setting. 
Initially, the ground state of the system with the above potential barrier $V(x,y)$ inside either type of external traps is identified.
Experimentally, this implies loading the relevant background into the overall potential landscape in order to craft the kink structure. 
Afterwards, $V(x,y)$ is suddenly removed and the system is monitored in the course of the evolution for inspecting the dynamical stability of the kink configuration.

Density snapshots following the aforementioned protocol are exemplarily illustrated in Fig.~\ref{Fig:Kink_engineering}(a) (Fig.~\ref{Fig:Kink_engineering}(b)) for the 2D eGPE (CQ) setting.
For these models, the interaction strengths are identical to the ones used in the main text and the 2D box length is $L = 400$.
As it can be seen, in the case of a 2D box potential the kink remains intact throughout the evolution in both models, see Fig.~\ref{Fig:Kink_engineering}(a) and Fig.~\ref{Fig:Kink_engineering}(b) respectively.
Shortly after the barrier ($V(x,y)$) release, the kink adjusts its width and amplitude towards its expected asymptotic values by emitting radiation traveling across its finite background, see the undulated tail in Fig.~\ref{Fig:Kink_engineering}(a) for the 2D eGPE and in Fig.~\ref{Fig:Kink_engineering}(b) for the CQ.
These modulations keep propagating during the dynamics
(eventually being transported outside the observation 
window since the latter is much smaller than the
overall box length), while the kink settles at the origin retaining its expected characteristics.

A similar dynamical process occurs in the 3D eGPE featuring a strong harmonic confinement along the $z$-direction
(while being homogeneous along the $y$-one), see Fig.~\ref{Fig:Kink_engineering}(c).  
Also here, the overall dynamics reveals stability of the kink state (with the appropriate quasi-2D asymptotics) suffering perturbations in the form of emitted  radiation attributed to the presence of  transverse excitations, see Fig.~\ref{Fig:Kink_engineering}(c). As such, we confirm that also in the presence of this $z$-direction transverse harmonic confinement the kink retains its dynamical stability,
including too transverse modulations along the $y$-direction.

\section{Three-dimensional attractive mixture}  \label{Supp:3D_eGPE}

The three-dimensional setting discussed in the main text corresponds to an attractive bosonic mixture, where the intercomponent attraction slightly exceeds the intracomponent repulsion.
Namely, $a_{12} +\sqrt{a_{11} a_{22}} \lesssim 0$, with $a_{ij}$ denoting the relevant 3D scattering lengths that are tunable with the Feshbach resonance technique~\cite{semeghini2018self}. 
In this interaction regime, it is theoretically predicted~\cite{petrov2015quantum,luo2021new} and experimentally observed~\cite{cabrera2018quantum,semeghini2018self} that the bosonic mixture hosts many-body self-bound states, known as quantum droplets.
The beyond mean-field first-order LHY correction is proven to be crucial for sustaining such structures, minimizing the underlying energy functional of the mixture.
The resulting extended mean-field equations (eGPEs) then feature competing nonlinearities, expressed through a cubic and a quartic term~\cite{petrov2015quantum}.

When the ratio between the density of the components satisfies $n_1/n_2=\sqrt{a_{22}/a_{11}}$, 
the two coupled extended mean-field equations reduce to a single effective one~\cite{semeghini2018self}, describing an attractive bosonic mixture.
The dimensional form of the eGPE reads~\cite{semeghini2018self}

\begin{gather}
    i \hbar \frac{\partial \Psi}{\partial t} = - \frac{\hbar^2}{2m} \nabla^2 \Psi  + \frac{8\pi \hbar^2 \delta a}{m}  \frac{\sqrt{a_{22}/a_{11}}}{(1+\sqrt{a_{22}/a_{11}})^2} \abs{\Psi}^2 \Psi \nonumber \\
    + \frac{128 \sqrt{\pi} \hbar^2}{3m}  (a_{11} a_{22})^{5/4} \abs{\Psi}^3 \Psi.
    \label{Eq:3D_eGPE_dimensionfull}
\end{gather}
Here, $m$ is the mass of the atoms, $\Psi$ is the 3D wave function describing the reduced single component, and the averaged mean-field interactions
 $\delta a =a_{12} +\sqrt{a_{11} a_{22}}$.

To render Eq. \eqref{Eq:3D_eGPE_dimensionfull} dimensionless, the following transformations are employed,
\begin{equation}
    t = \tilde{t}  \frac{m L^2}{\hbar}, \quad r=\tilde{r} L, \quad \Psi= \tilde{\Psi} L^{-3/2},
    \label{Eq:Dimensionalization}
\end{equation}
where e.g. $L\simeq  1.6 \, \mu m$ is the length of the box along the $z$ direction.
With the above units, the 3D eGPE becomes dimensionless,

\begin{equation}
    i \frac{\partial \Psi}{\partial t} = -\frac{1}{2} \nabla^2 \Psi
    +g_1 \abs{\Psi}^2 \Psi +g_2 \abs{\Psi}^3 \Psi.
    \label{Eq:3D_eGPE_dimensionless}
\end{equation}
The $\sim$ notation has been dropped for simplicity. Moreover, the dimensionless interaction strengths $g_1, g_2$ in these units read

\begin{subequations}
\begin{gather}
    g_1 =  \frac{8 \pi \delta a \sqrt{a_{22}/a_{11}}}{ L (1+\sqrt{a_{22}/a_{11}})^2 },
    \label{Eq:3D_interaction_strength} \\
    g_2 = \frac{128 \sqrt{\pi}}{3} \left(  a_{11} a_{22}  \right)^{5/4} L^{-5/2}.
\end{gather}    
\end{subequations}

Rescaling the wavefunction as $\Psi = \Psi' g_2^{-1/3}$, we absorb the prefactor in front of the quartic term and we recover Eq.~(1c) in the main text.
For the two hyperfine states of $^{39}$K~\cite{semeghini2018self,cabrera2018quantum,cheiney2018bright},
it was shown that droplets can form when $\delta a$ is tuned to small negative values by means of Fano-Feshbach resonances~\cite{chin2010feshbach}. 
Relevant timescales correspond to $mL^2/\hbar  \approx 1.6 \, ms$.

\section{Two-dimensional droplet setting}   \label{Supp:2D_eGPE}

Here, we elaborate on the 2D eGPE supporting quantum droplets, its dimensionless form, and the connection of the interaction strengths with the experimentally tunable 3D scattering lengths.
Quantum droplets can also form in two dimensions~\cite{petrov2016ultradilute}, where the logarithmic nonlinearity incorporates both the LHY term and the mean-field contributions.
Similar to 3D, the corresponding coupled system of eGPEs can be reduced to an effective single-component one~\cite{Otajonov_variational_2020}, which reads

\begin{equation}
i \hbar \frac{\partial \Psi}{\partial t} = -\frac{\hbar^2}{2m} \nabla^2 \Psi +\frac{\hbar^2}{m} g |\Psi|^2 \Psi \ln \left(  \frac{|\Psi|^2}{n_0 \sqrt{e}} \right).
\label{Eq:2D_eGPE_supp}
\end{equation}
The dimensionless interaction strength $g$ is related to the 2D scattering lengths through the expression~\cite{petrov2016ultradilute,Otajonov_variational_2020},
\begin{subequations}
\begin{gather}
g = \frac{4 \pi}{\ln \left(  \frac{a^{(2D)}_{12}\sqrt{a^{(2D)}_{11}a^{(2D)}_{22}}}{[a^{(2D)}_{11}]^2 \Delta }  \right)   \ln  \left(   \frac{a^{(2D)}_{12}\sqrt{a^{(2D)}_{11}a^{(2D)}_{22}}}{[a^{(2D)}_{22}]^2\Delta }    \right)},
\label{Eq:Interaction_strength} \\
\rm{where}~~
\Delta = \exp\left\{ -\frac{\ln^2 \left( \frac{a^{(2D)}_{22}}{a^{(2D)}_{11}}  \right)}{2 \ln \left(   \frac{[a^{(2D)}_{12}]^2}{a^{(2D)}_{11} a^{(2D)}_{22}}  \right)}  \right \}. \label{Eq:delta_parameter}
\end{gather} 
\end{subequations}
In these expressions, $a^{(2D)}_{\sigma \sigma'}$ refer to the 2D scattering lengths within ($\sigma=\sigma'$) or between ($\sigma \neq \sigma'$) the components. 
Moreover, 
$n_0 = \frac{e^{-2 \gamma -3/2}}{\pi a^{(2D)}_{12} \sqrt{a^{(2D)}_{11} a^{(2D)}_{22}}} \Delta \sqrt{
\frac{4 \pi}{g}}$ is the droplet equilibrium density in the thermodynamic limit~\cite{petrov2016ultradilute} and  $\gamma \approx 0.577$ is the Euler-Mascheroni constant~\cite{abramowitz_handbook_1988}.

In the case of a tightly confined mixture in the $z$-direction due to the presence of a harmonic trap with oscillator length $a_{\perp}=\sqrt{\hbar/(m \omega_{\perp})}$,
the scattering lengths are connected to their 3D counterparts, $a_{\sigma \sigma'}$ by~\cite{Petrov_interatomic_2001,Petrov_Bose_2000}
\begin{equation}
a^{(2D)}_{\sigma \sigma'} = 2e^{-\gamma} a_{\perp} \sqrt{\frac{\pi}{0.9}} \exp  \left \{   -\frac{a_{\perp} \sqrt{\pi}}{a_{\sigma \sigma'} \sqrt{2}}  \right  \}.
\label{Eq:Scattering_length_mapping}
\end{equation}
Below, we measure time and length with respect to $m/(g n_0 \hbar \sqrt{e})$ and $\sqrt{g n_0 \sqrt{e}}^{-1}$ respectively~\cite{Jalm_dynamical_2019} and normalize the 2D wave function in terms of the total particle number, $\int |\Psi|^2~dx dy =gN=g(N_1+ N_2)$. 
In this sense the corresponding dimensionless eGPE is retrieved,
\begin{equation}
i \frac{\partial \Psi}{\partial t} = -\frac{1}{2} \nabla^2 \Psi +  |\Psi|^2 \Psi \ln\left(|\Psi|^2\right).
\label{Eq:dimensionless_eGPE}
\end{equation}

\section{Cubic-quintic model} \label{Supp:CQ}

Below we elaborate on the dimensionalization of the cubic-quintic equation in the context of non-linear optical media discussed in the main text [Eq. (1b)].
Particularly, the relevant model considers light beams passing through nonlinear media, being characterized by high-order refractive index 
prefactors (stemming from an expansion of
the refractive index as a function of
the light intensity). 
In this context, the propagation of the electric field envelope $E=E(x,y,z)$ along the axial direction $z$ can be described by the following cubic-quintic equation~\cite{Reyna_observation_2020,Filho_robust_2013},

\begin{equation}
    i \frac{\partial E}{\partial z} = -\frac{1}{2k} \nabla^2 E -2 k n_2 c \epsilon_0 \abs{E}^2E-4kn_4 c^2 \epsilon_0^2 \abs{E}^4 E.
    \label{Eq:CQ_nonlinear_optics}
\end{equation}
Here, $k$ is the wavenumber, $n_0$ refers to the linear refractive index, and $n_2$, $n_4$ are the higher-order nonlinear refractive indices.
Moreover, $\epsilon_0$ and $c$ are the vacuum permittivity and speed of light respectively. 
The nabla operator describes the dispersion of the beam along the transversal directions, $x$ and $y$.

To cast the above equation in dimensionless form, we employ the coordinate transformations, $ \{x,y,z  \} = r_0 \{\tilde{x},\tilde{y},\tilde{z}  \}$, and $E= E_0 \tilde{E}$ where,
\begin{equation}
    r_0=k^{-1}, \quad E_0 = \frac{1}{\sqrt{2 n_2 c \epsilon_0}}.
    \label{Eq:Dimensionalization_CQ}
\end{equation}
The CQ subsequently reads,
\begin{equation}
    i \frac{\partial E}{\partial z} = -\frac{1}{2} \nabla^2 E -  \abs{E}^2 E +g_{CQ} \abs{E}^4 E,
    \label{Eq:CQ_dimensionless}
\end{equation}
where $g_{CQ}= -n_4/n_2^2$. For liquid carbon disulfide $\rm{CS}_2$ employed in Refs.~\cite{Reyna_observation_2020,Filho_robust_2013}, $g_{CQ} =450$.


\end{document}